\documentclass{moriond}

\def\be{\begin{equation}}
\def\ee{\end{equation}}
\def\bea{\begin{eqnarray}}
\def\eea{\end{eqnarray}}

\def\gsim{\mathrel{\rlap{\lower4pt\hbox{\hskip1pt$\sim$}}
    \raise1pt\hbox{$>$}}}         
\def\lsim{\mathrel{\rlap{\lower4pt\hbox{\hskip1pt$\sim$}}
    \raise1pt\hbox{$<$}}}         

\newcommand{\Photo}{\includegraphics[height=35mm]{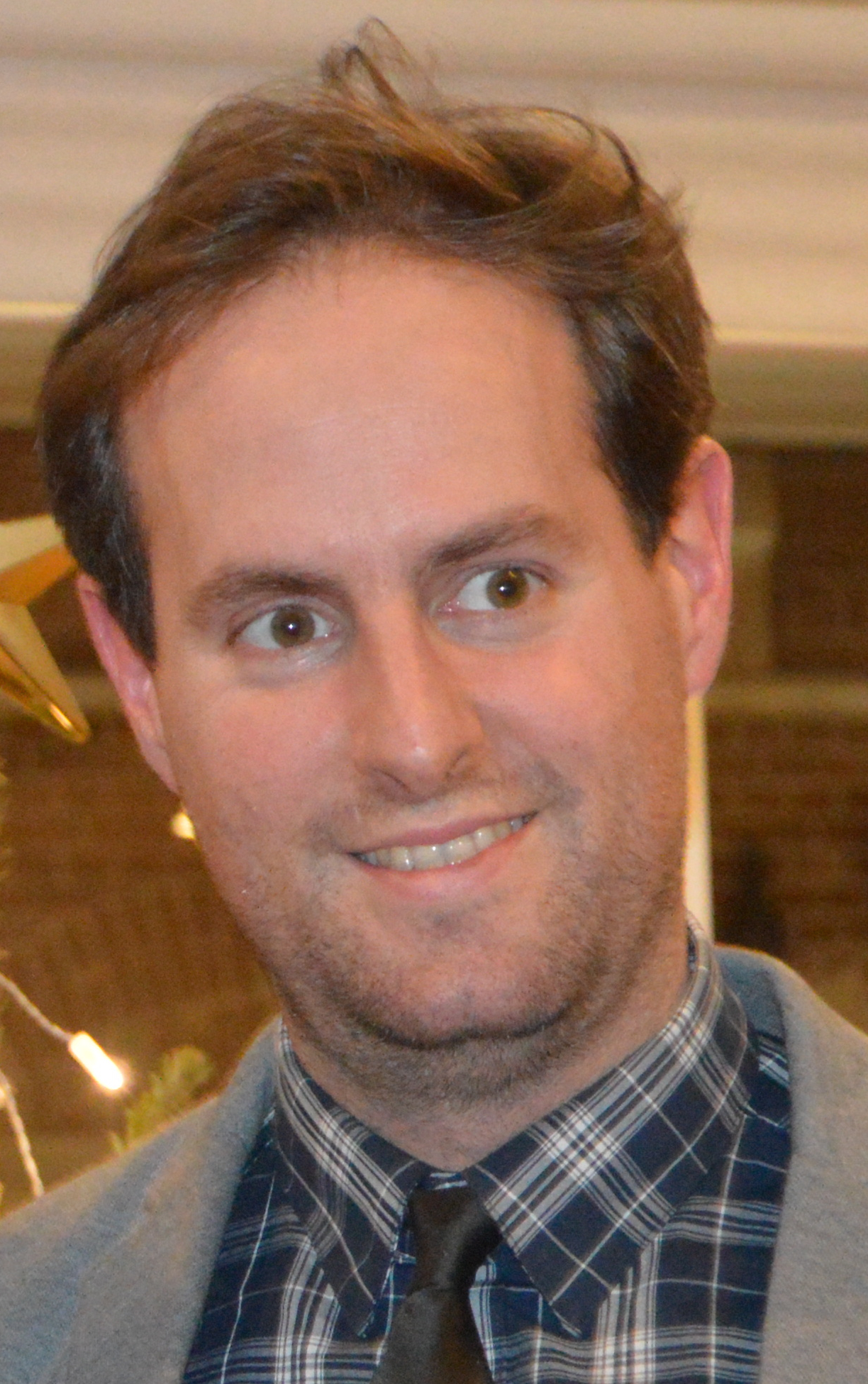}}

\begin{document}
\vspace*{4cm}
\title{PROGRESS IN THE NNPDF GLOBAL ANALYSES\\ OF PROTON STRUCTURE}

\author{ Juan Rojo }

\address{ Department of Physics and Astronomy, Vrije Universiteit, 1081 HV Amsterdam.\\
Nikhef Theory Group, Science Park 105, 1098 XG Amsterdam, The Netherlands.}

\maketitle\abstracts{
  I present recent progress in the NNPDF global analyses of parton
  distributions (PDFs) focusing on developments contributing
  to its new upcoming release: NNPDF4.0.
  The NNPDF4.0 determination represents unprecedented  progress in three main directions:
  {\it i)} the systematic inclusion of LHC Run II data at $\sqrt{s}=13$ TeV and of
  new processes from dijets to single top distributions, {\it ii)} the deployment of
  state-of-the-art machine learning
  algorithms, from automated hyperparameter optimisation to stochastic gradient descent
  training; and {\it iii)} the complete statistical validation of PDF uncertainties,
  both in the data and extrapolation regions, by means of closure and future tests.
  Other methodological
  improvements in NNPDF4.0 include strict PDF positivity, integrability constraints at small-$x$,
  and deuteron and heavy nuclear corrections.
  I present representative results from NNPDF4.0 and 
 assess its  impact  on open issues such as the light anti-quark asymmetry  and
  the  charm content of protons.
}

\noindent
{\bf The road towards NNPDF4.0.}
Since the release of the NNPDF3.1 global analysis~\cite{Ball:2017nwa} in 2017, a significant amount of work
has been invested in laying the groundwork for the subsequent major update, dubbed NNPDF4.0
and now close to completion.
Progress towards NNPDF4.0 has taken place along several complementary directions,
such as addition of new datasets, many of them from Run II of the LHC, and altogether new types of processes,
the implementation of powerful machine learning fitting tools,
novel strategies for the statistical validation of PDF uncertainties, 
and the refined implementation of theoretical constraints that restrict the possible shapes
that the PDFs are allowed to take.

\begin{figure}[t]
  \begin{center}
     \includegraphics[width=0.49\linewidth]{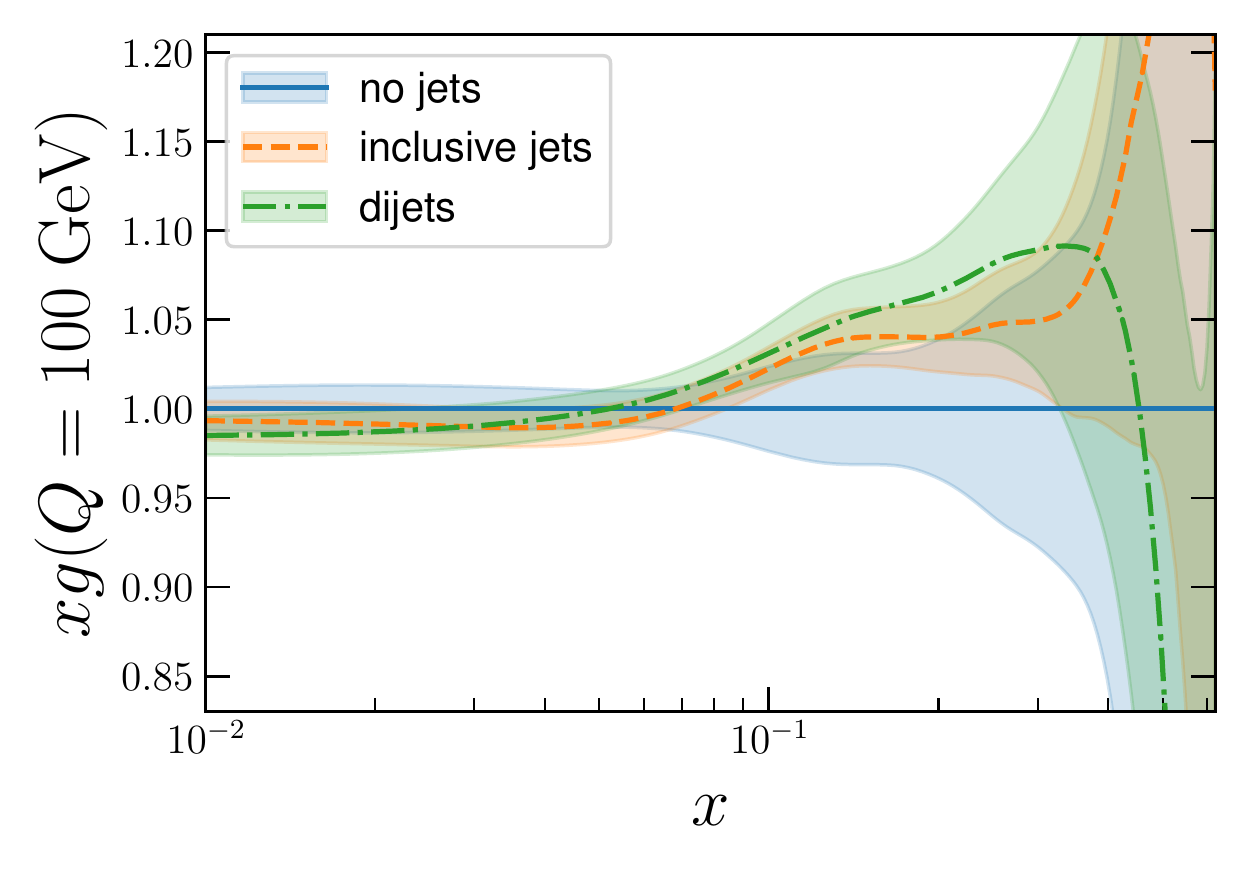}
     \includegraphics[width=0.49\linewidth]{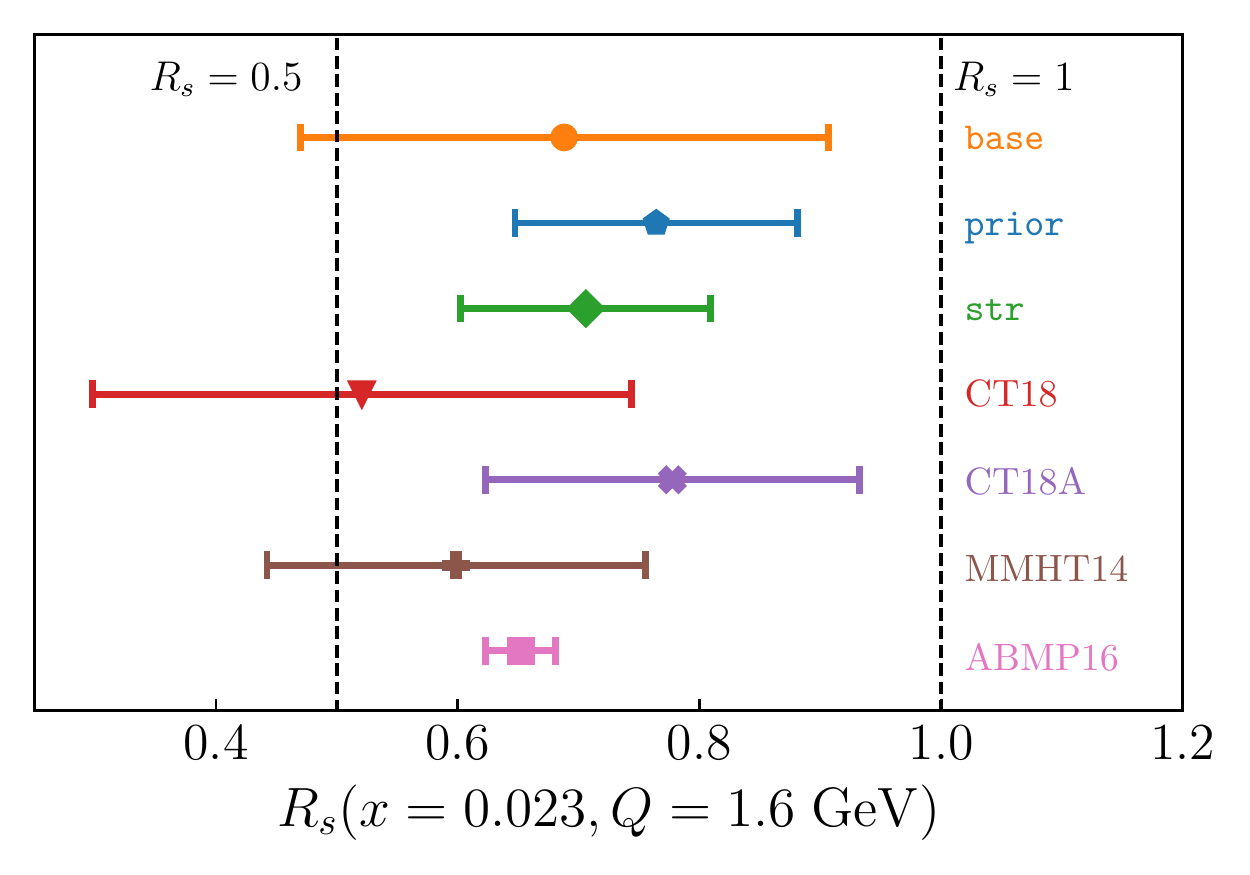}
    \caption{\small
      Left: comparative impact on the gluon PDF of available inclusive jets and dijet
      cross-sections,
      where here the baseline PDF fit does not include any jet data.
      Right: the strangeness ratio $R_s(x,Q)$ evaluated at $x=0.023$ and $Q=1.6$ GeV,
      comparing the predictions of various global fits with
      the {\tt NNPDF3.1\_str}
      analysis which includes the constraints from the NOMAD neutrino
      cross-sections.
  \label{fig:newdata} }
\end{center}
\end{figure}

New groups of processes added for the first time in NNPDF4.0 include dijet
cross-sections, single top quark distributions, direct photon production,
and $W$ boson production in association with jets, among several others.
As an illustration of the impact of these new processes, the left panel Fig.~\ref{fig:newdata} demonstrates~\cite{AbdulKhalek:2020jut} that the
constraints on the gluon PDF from Run I inclusive jet and dijet
cross-sections is qualitatively consistent when added on top
of a baseline NNPDF3.1-like fit that does not include any jet data.
Furthermore, all available 7 and 8 TeV dijet cross-sections are successfully described
by NNLO QCD theory once included in the fit, without the need to introduce tailored
decorrelation models as is sometimes the case with inclusive jets.

Another data-related study on the road towards NNPDF4.0 was the
reappraisal of the proton strangeness content based on the
combination of all relevant experimental inputs~\cite{Faura:2020oom}, among them
the NOMAD dimuon neutrino-induced cross-sections which provide
significant constraints.
Interestingly, a satisfactory joint description of all datasets is found,
with no evidence for tensions between groups of processes.
 As indicated by Fig.~\ref{fig:newdata}, which displays the
 ratio $R_s(x,Q)=(s+\bar{s})/(\bar{u}+\bar{d})$ at $x=0.023$ and $Q=1.6$ GeV,
 a global analysis of all strangeness-sensitive measurements
 favors a moderately suppressed strangeness PDF,
 and disfavors scenarios with either a heavily suppressed ($R_s\lsim 0.5$)
 of a symmetric ($R_s\simeq 1$) strange sea.
 
Concerning the improved treatment of theoretical constraints,
NNPDF4.0 accounts for the strict positivity of $\overline{\rm MS}$ PDFs~\cite{Candido:2020yat},
the  integrability of the non-singlet quark combinations $T_3$ and $T_8$ at small-$x$,
and the uncertainties associated to deuteron and heavy nuclear effects for data involving
nuclear targets.
The latter, relevant in particular for the description of the NuTeV and CHORUS
neutrino structure functions, is implemented using the method from~\cite{Ball:2018twp}
with the nNNPDF2.0 global nPDF fit~\cite{AbdulKhalek:2020yuc} as input.
This choice is particularly suitable here, given that the fitting
methodology~\cite{AbdulKhalek:2019mzd}
of nNNPDF2.0 is consistent with that used in the NNPDF4.0 analysis.

From the point of view of
machine learning tools, NNPDF4.0 benefits from a suite of state-of-the-art
stochastic gradient descent methods for neural network training,
combined with the automated optimisation of  model hyperparameters such as the
network architecture~\cite{Carrazza:2019mzf}.
These improvements have resulted into a dramatic speed-up of the per-replica running
time of up to a factor 10, as well as a reduction of the PDF uncertainties by removing
minimisation inefficiencies present in genetic algorithms.
The statistical interpretation of PDF uncertainties is then carefully validated
by means of closure tests~\cite{Ball:2014uwa,Forte:2020yip}, where PDFs are fitted to pseudo-data generated
with a known underlying law, and future tests~\cite{Cruz-Martinez:2021rgy}, which verify
the forecasting performance
of the new methodology to predict novel datasets with different kinematic
coverage.\\[-0.3cm]

\begin{figure}[t]
  \begin{center}
    \includegraphics[width=0.49\linewidth]{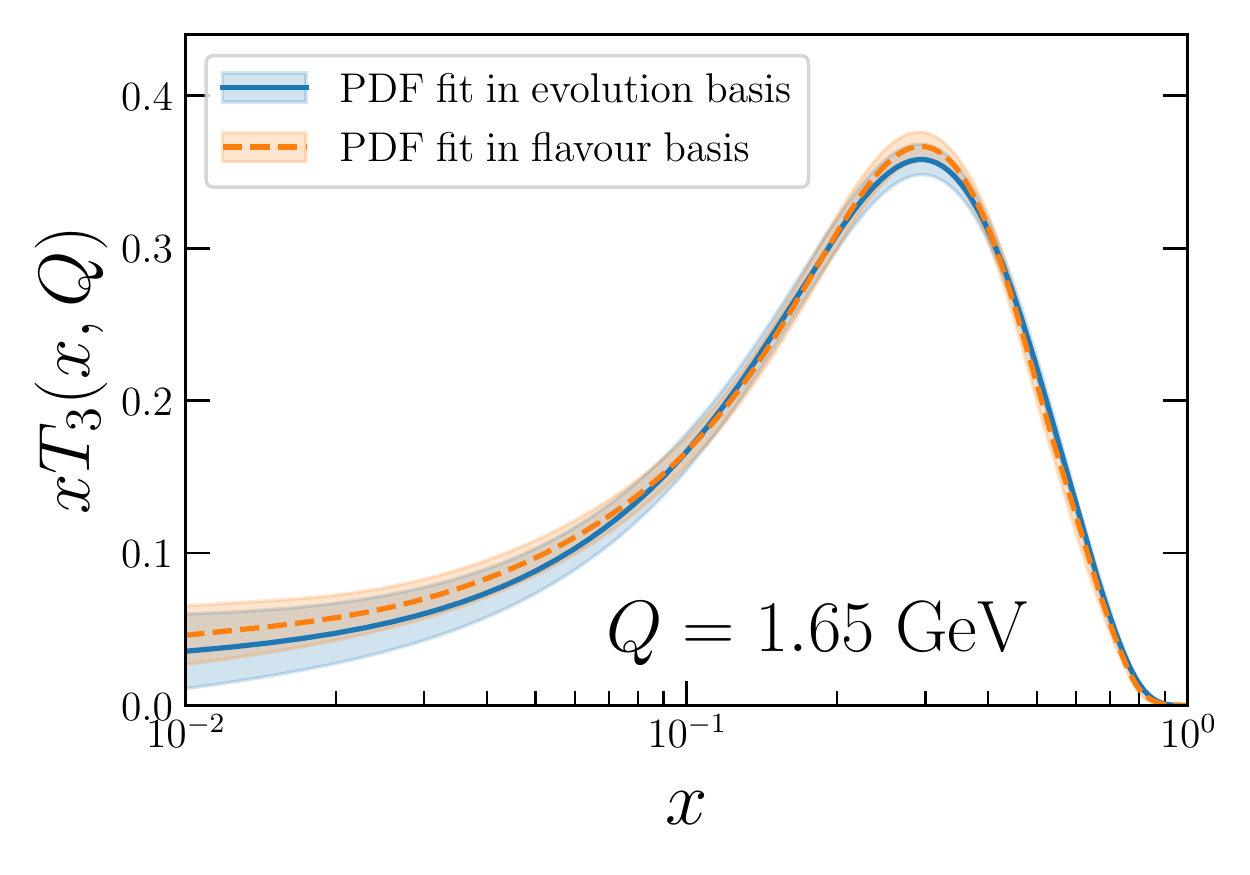}
    \includegraphics[width=0.49\linewidth]{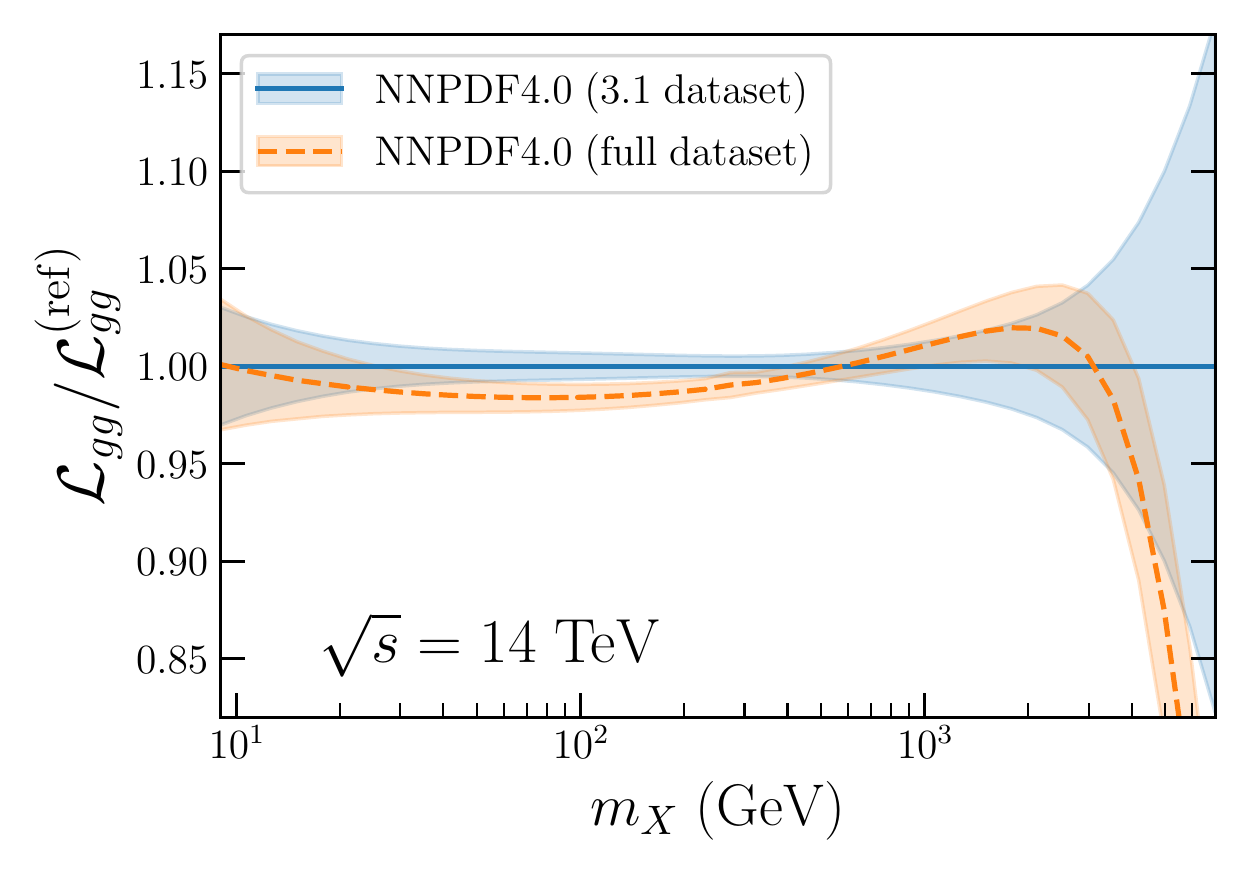}
    \caption{\small Left: comparison of the nonsinglet quark triplet, $T_3=u+\bar{u}-d-\bar{d}$,
      in NNPDF4.0 fits performed in either the evolution or the flavour parametrisation basis.
      Right: the gluon-gluon luminosity at $\sqrt{s}=14$ TeV as a function of $m_X$,
      displaying the impact
      of the new data added in NNPDF4.0.
  \label{fig:nnpdf40} }
\end{center}
\end{figure}

\noindent
{\bf A first look at NNPDF4.0.}
A highly non-trivial self-consistency test of the new fitting methodology
is that one is able to carry out PDF fits either in the evolution basis
(where the NNs parametrise $\Sigma, T_3, T_8, V$, and so on) or in the flavour
basis (where instead they parametrise $u,\bar{u}, d,\bar{d}$, and so on) finding
fully consistent results.
This feature is highlighted in Fig.~\ref{fig:nnpdf40}, which compares
the output of the  nonsinglet quark triplet, $T_3=u+\bar{u}-d-\bar{d}$, at
the initial scale $Q=1.65$ GeV obtained
in the two parametrisation bases: excellent agreement is obtained, illustrating
the robust basis independence of the NNPDF4.0 global analysis.

The right panel of Fig.~\ref{fig:nnpdf40} quantifies the impact of the new
data added in NNPDF4.0 on the gluon-gluon partonic luminosity at $\sqrt{s}=14$ TeV as a function of
invariant mass $m_X$.
In general, one finds agreement at the one-sigma level, with the new data (driven by  dijet
cross-sections) modifying the
overall shape of $\mathcal{L}_{gg}$
with a slight suppression for $m_X\simeq 100$ GeV, an enhancement starting
at 1 TeV, and then again a suppression from masses of 4 TeV onwards.
The new data also leads to a reduction of the PDF errors in $\mathcal{L}_{gg}$
in the region with $m_X\gsim 500$ GeV.

\begin{figure}[t]
  \begin{center}
    \includegraphics[width=0.49\linewidth]{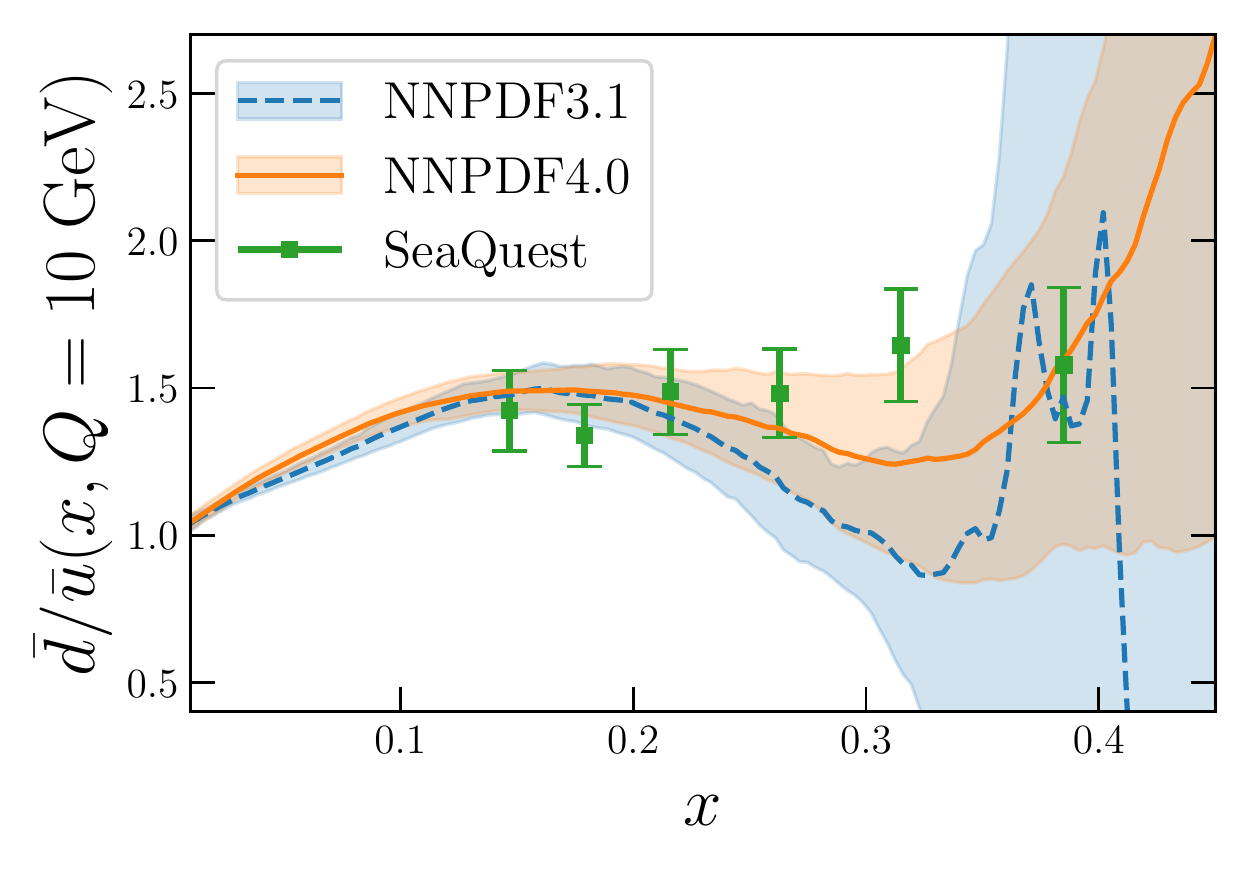}
    \includegraphics[width=0.49\linewidth]{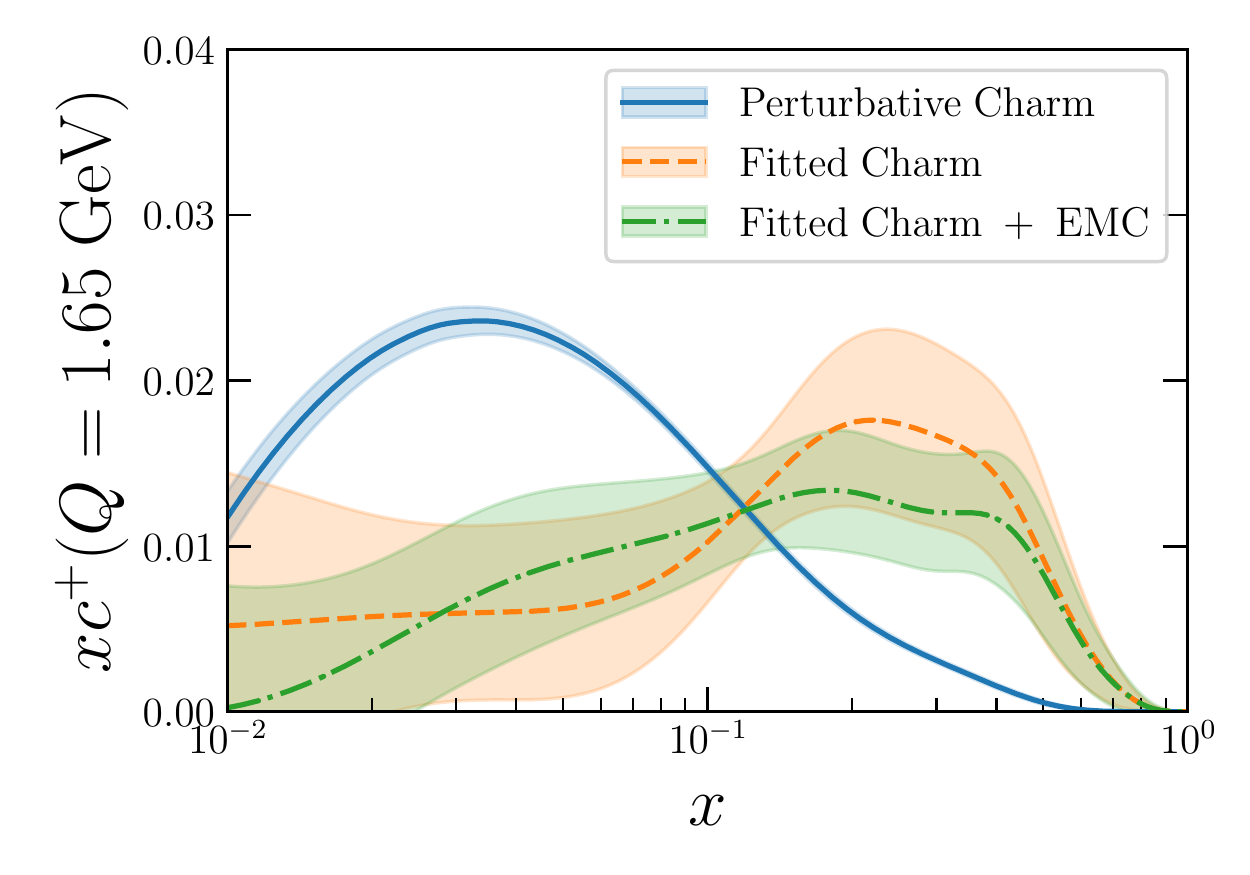}
    \caption{\small Left: comparison of the NNPDF3.1 and NNPDF4.0 predictions
      for the light anti-quark PDF ratio $\bar{d}/\bar{u}$ at $Q=10$ GeV
      with the SeaQuest results.
      Right: the charm PDF at $Q=1.65$ GeV in the
      NNPDF4.0 fits with perturbative charm and with fitted charm (with and without EMC charm data)
  \label{fig:charm_content} }
\end{center}
\end{figure}

The novel NNPDF4.0 analysis also makes possible a high-precision assessment
of many important open questions releated to the non-perturbative QCD nature
of proton structure.
To illustrate this potential, Fig.~\ref{fig:charm_content}  displays the
NNPDF3.1 and NNPDF4.0 predictions
for the light anti-quark PDF ratio, $\bar{d}(x,Q)/\bar{u}(x,Q)$ at $Q=10$ GeV.
This partonic ratio has also been recently evaluated by the SeaQuest experiment~\cite{Dove:2021ejl},
following
a model-dependent deconvolution from their cross-section measurements.
We observe that both the NNPDF3.1 and NNPDF4.0 predictions
are in good agreement with the SeaQuest determinations of the light quark
sea asymmetry.

Then the right panel of  Fig.~\ref{fig:charm_content}  displays
the charm PDF at $Q=1.65$ GeV in the
NNPDF4.0 fits.
We provide results obtained both with perturbative charm and with fitted charm~\cite{Ball:2016neh},
in the latter
case with and without the EMC charm structure functions included in the fit.
This comparison highlights how current data favours a valence-like structure for the charm
PDF at low-scales, which in turn is consistent with the hypothesis of an intrinsic charm component
in the proton wave function.
We note that the addition or not of the EMC data does not modify in a significant manner the resulting charm
PDF, highlighting how the dominant constraints on $c(x,Q_0)$ arise from other datasets such as forward
$W,Z$ production by LHCb.\\[-0.3cm]

\noindent
{\bf The road ahead.}
The global NNPDF4.0 determination achieves an unprecedented accuracy in a broad kinematic range,
thanks to its extensive dataset combined with deep-learning optimisation models.
Its faithfulness in representing PDF uncertainties is validated in detail by closure tests, future tests, and parametrisation basis independence studies.
Given the current level of PDF uncertainties obtained with NNPDF4.0,
it becomes of paramount importance to  continue the efforts towards the inclusion in the
fit of various types of theoretical uncertainties, such as those arising from missing higher orders~\cite{AbdulKhalek:2019ihb} and
from the SM parameters such as $\alpha_s(m_Z)$~\cite{Ball:2018iqk}, and well as to account for higher order terms
in the QCD (eventually up to N3LO) and EW perturbative expansions, in the latter case using
recently developed tools such as {\tt PineAPPL}~\cite{Carrazza:2020gss}.\\[-0.3cm]

\noindent
    {\bf An open-source machine learning fitting framework}.
    The NNPDF machine learning fitting framework will be publicly released, fully open source,
    together with an extensive documentation and user-friendly examples.
We believe that its availability will provide a useful resource for the community in cases where
one aims to apply cutting-edge machine learning tools to the model-independent parametrisation
of non-perturbative QCD quantities, as well as to
related applications both within and beyond~\cite{Roest:2020kqy} high-energy physics.

{\small
\section*{References}


\begin{thebibliography}{10}

\bibitem{Ball:2017nwa}
{\bf NNPDF} Collaboration, R.~D. Ball et~al., {\it {Parton distributions from
  high-precision collider data}},  {\em Eur. Phys. J.} {\bf C77} (2017), no.~10
  663, [\href{http://arxiv.org/abs/1706.00428}{{\tt arXiv:1706.00428}}].

\bibitem{AbdulKhalek:2020jut}
R.~Abdul~Khalek et~al., {\it {Phenomenology of NNLO jet production at the LHC
  and its impact on parton distributions}},  {\em Eur. Phys. J. C} {\bf 80}
  (2020), no.~8 797, [\href{http://arxiv.org/abs/2005.11327}{{\tt
  arXiv:2005.11327}}].

\bibitem{Faura:2020oom}
F.~Faura, S.~Iranipour, E.~R. Nocera, J.~Rojo, and M.~Ubiali, {\it {The
  Strangest Proton?}},  {\em Eur. Phys. J. C} {\bf 80} (2020), no.~12 1168,
  [\href{http://arxiv.org/abs/2009.00014}{{\tt arXiv:2009.00014}}].

\bibitem{Candido:2020yat}
A.~Candido, S.~Forte, and F.~Hekhorn, {\it {Can $ \overline{\mathrm{MS}} $
  parton distributions be negative?}},  {\em JHEP} {\bf 11} (2020) 129,
  [\href{http://arxiv.org/abs/2006.07377}{{\tt arXiv:2006.07377}}].

\bibitem{Ball:2018twp}
{\bf NNPDF} Collaboration, R.~D. Ball, E.~R. Nocera, and R.~L. Pearson, {\it
  {Nuclear Uncertainties in the Determination of Proton PDFs}},  {\em Eur.
  Phys. J.} {\bf C79} (2019), no.~3 282,
  [\href{http://arxiv.org/abs/1812.09074}{{\tt arXiv:1812.09074}}].

\bibitem{AbdulKhalek:2020yuc}
R.~Abdul~Khalek, J.~J. Ethier, J.~Rojo, and G.~van Weelden, {\it {nNNPDF2.0:
  quark flavor separation in nuclei from LHC data}},  {\em JHEP} {\bf 09}
  (2020) 183, [\href{http://arxiv.org/abs/2006.14629}{{\tt arXiv:2006.14629}}].

\bibitem{AbdulKhalek:2019mzd}
{\bf NNPDF} Collaboration, R.~Abdul~Khalek, J.~J. Ethier, and J.~Rojo, {\it
  {Nuclear parton distributions from lepton-nucleus scattering and the impact
  of an electron-ion collider}},  {\em Eur. Phys. J. C} {\bf 79} (2019), no.~6
  471, [\href{http://arxiv.org/abs/1904.00018}{{\tt arXiv:1904.00018}}].

\bibitem{Carrazza:2019mzf}
S.~Carrazza and J.~Cruz-Martinez, {\it {Towards a new generation of parton
  densities with deep learning models}},  {\em Eur. Phys. J. C} {\bf 79}
  (2019), no.~8 676, [\href{http://arxiv.org/abs/1907.05075}{{\tt
  arXiv:1907.05075}}].

\bibitem{Ball:2014uwa}
{\bf NNPDF} Collaboration, R.~D. Ball et~al., {\it {Parton distributions for
  the LHC Run II}},  {\em JHEP} {\bf 04} (2015) 040,
  [\href{http://arxiv.org/abs/1410.8849}{{\tt arXiv:1410.8849}}].

\bibitem{Forte:2020yip}
S.~Forte and S.~Carrazza, {\it {Parton distribution functions}},
  \href{http://arxiv.org/abs/2008.12305}{{\tt arXiv:2008.12305}}.

\bibitem{Cruz-Martinez:2021rgy}
J.~Cruz-Martinez, S.~Forte, and E.~R. Nocera, {\it {Future tests of parton
  distributions}},  in {\em {60th Cracow School of Theoretical Physics}:
  {Panorama of Hadronic Physics}}, 3, 2021.
\newblock \href{http://arxiv.org/abs/2103.08606}{{\tt arXiv:2103.08606}}.

\bibitem{Dove:2021ejl}
{\bf SeaQuest} Collaboration, J.~Dove et~al., {\it {The asymmetry of antimatter
  in the proton}},  {\em Nature} {\bf 590} (2021), no.~7847 561--565,
  [\href{http://arxiv.org/abs/2103.04024}{{\tt arXiv:2103.04024}}].

\bibitem{Ball:2016neh}
{\bf NNPDF} Collaboration, R.~D. Ball, V.~Bertone, M.~Bonvini, S.~Carrazza,
  S.~Forte, A.~Guffanti, N.~P. Hartland, J.~Rojo, and L.~Rottoli, {\it {A
  Determination of the Charm Content of the Proton}},  {\em Eur. Phys. J.} {\bf
  C76} (2016), no.~11 647, [\href{http://arxiv.org/abs/1605.06515}{{\tt
  arXiv:1605.06515}}].

\bibitem{AbdulKhalek:2019ihb}
{\bf NNPDF} Collaboration, R.~Abdul~Khalek et~al., {\it {Parton Distributions
  with Theory Uncertainties: General Formalism and First Phenomenological
  Studies}},  {\em Eur. Phys. J. C} {\bf 79} (2019), no.~11 931,
  [\href{http://arxiv.org/abs/1906.10698}{{\tt arXiv:1906.10698}}].

\bibitem{Ball:2018iqk}
{\bf NNPDF} Collaboration, R.~D. Ball, S.~Carrazza, L.~Del~Debbio, S.~Forte,
  Z.~Kassabov, J.~Rojo, E.~Slade, and M.~Ubiali, {\it {Precision determination
  of the strong coupling constant within a global PDF analysis}},  {\em Eur.
  Phys. J.} {\bf C78} (2018), no.~5 408,
  [\href{http://arxiv.org/abs/1802.03398}{{\tt arXiv:1802.03398}}].

\bibitem{Carrazza:2020gss}
S.~Carrazza, E.~R. Nocera, C.~Schwan, and M.~Zaro, {\it {PineAPPL: combining EW
  and QCD corrections for fast evaluation of LHC processes}},  {\em JHEP} {\bf
  12} (2020) 108, [\href{http://arxiv.org/abs/2008.12789}{{\tt
  arXiv:2008.12789}}].

\bibitem{Roest:2020kqy}
L.~I. Roest, S.~E. van Heijst, L.~Maduro, J.~Rojo, and S.~Conesa-Boj, {\it
  {Charting the low-loss region in Electron Energy Loss Spectroscopy with
  machine learning}},  {\em Ultramicroscopy} {\bf 222} (2021) 113202,
  [\href{http://arxiv.org/abs/2009.05050}{{\tt arXiv:2009.05050}}].

\end{thebibliography}
\providecommand{\href}[2]{#2}\begingroup\raggedright\endgroup

}

\end{document}